\documentclass[final,aps,prb,twocolumn,superscriptaddress,showpacs]{revtex4-1}

\usepackage{graphicx}
\usepackage{dcolumn}
\usepackage{amsmath,bm}
\usepackage{color}%
\usepackage{multirow}
\usepackage{array} 
\usepackage{esvect}
\usepackage{braket}
\usepackage{makecell}
\usepackage[utf8]{inputenc}
\usepackage{kotex}
\usepackage{xr}
\usepackage{soul,xcolor}
\setstcolor{red} 

\usepackage{xr}

\makeatletter
\newcommand*{\addFileDependency}[1]{
  \typeout{(#1)}
  \@addtofilelist{#1}
  \IfFileExists{#1}{}{\typeout{No file #1.}}
}
\makeatother

\newcommand*{\myexternaldocument}[1]{%
    \externaldocument{#1}%
    \addFileDependency{#1.tex}%
    \addFileDependency{#1.aux}%
}

\myexternaldocument{SI}  

\begin{document}

\title{Scalable thru-hole epitaxy of GaN through self-adjusting $h$-BN masks via solution-processed 2D stacks} 

\author{Jongwoo Ha}
\affiliation{Department of Physics and Research Institute for Basic Sciences, Kyung Hee University, Seoul, 02447, Republic of Korea}

\author{Minah Choi}
\affiliation{Department of Chemistry and Research Institute for Basic Sciences, Kyung Hee University, Seoul, 02447, Republic of Korea}

\author{Jieun Yang}
\email{jey@khu.ac.kr}
\affiliation{Department of Chemistry and Research Institute for Basic Sciences, Kyung Hee University, Seoul, 02447, Republic of Korea}

\author{Chinkyo Kim}
\email{ckim@khu.ac.kr}
\affiliation{Department of Physics and Research Institute for Basic Sciences, Kyung Hee University, Seoul, 02447, Republic of Korea}
\affiliation{Department of Information Display, Kyung Hee University, Seoul, 02447, Republic of Korea}

\date{\today}
\begin{abstract}
Selective epitaxy on 2D-material masks is a promising pathway for achieving localized, defect-suppressed GaN growth, but conventional 2D transfer processes limit scalability and interface control. Here, we demonstrate a thru-hole epitaxy (THE) method that enables vertically connected and laterally overgrown GaN domains through a spin-coated, solution-processed stack of hexagonal boron nitride ($h$-BN) flakes. The disordered $h$-BN mask exhibits a self-adjusting structure during growth, which locally reconfigures to allow percolative precursor transport and coherent GaN nucleation beneath otherwise blocking layers. Comprehensive structural analyses using scanning electron microscopy, Raman mapping, and high-resolution transmission electron microscopy confirm both the presence of epitaxial GaN beneath the h-BN and suppression of threading dislocations. This strategy eliminates the need for patterned 2D mask transfers and demonstrates a scalable route to selective-area GaN growth on arbitrary substrates, relevant to future micro-LED and photonic integration platforms.
\end{abstract}
\maketitle

\section{Introduction}
Epitaxial lateral overgrowth (ELOG) has been widely employed to grow defect-suppressed heteroepitaxial films on substrates with large lattice mismatches, using lithographically patterned dielectric masks. Since the advent of two-dimensional (2D) materials, a different strategy—thru-hole epitaxy (THE)—has emerged to further improve defect suppression by leveraging 2D materials as growth masks, where the efficacy of dislocation reduction is closely tied to the size and spatial distribution of nucleation openings.

While ELOG typically relies on selective nucleation through lithographically defined mask openings,\cite{Jang-SR-8-4112,Lee-JAC-52-532,Kim-CGD-20-6198} or through discontinuous 2D material films,\cite{Zhang-ACSAMI-7-4504} THE operates via diffusion of precursor species through percolative pathways within 2D material layers, followed by nucleation directly on the underlying substrate.\cite{Jang-AMI-10-2201406} In conventional ELOG, nucleation occurs rapidly at explicitly exposed substrate regions, followed by lateral overgrowth across the mask. In contrast, THE is governed by precursor diffusion through nanoscale transport pathways—either intrinsic defects in continuous 2D films or interconnected networks formed between loosely stacked flakes—that allow atomic species to reach the substrate through the 2D layer.\cite{Lee-CGD-22-6995,Lee-AEM-26-2301654} As a result, nucleation in THE is spatially and kinetically constrained, leading to delayed nucleation events and the formation of more isolated initial domains. Although lateral overgrowth eventually proceeds after nucleation, the fundamental distinction lies in how the substrate is initially accessed.

Although thru-hole epitaxy (THE) has been successfully demonstrated using transferred 2D material layers, achieving greater scalability and process simplicity would benefit from directly coating three-dimensional (3D) substrates with solution-processed 2D material suspensions. Compared to transferred or directly grown 2D layers, solution-processed films naturally exhibit broader distributions in flake thickness and lateral size. Liquid-phase exfoliation typically produces highly polydisperse flake populations, while intercalation-based methods achieve narrower—but still non-uniform—distributions.\cite{Kim-ACSMaterials-2-382}

These morphological variations become particularly problematic when the flake thickness exceeds a few atomic layers. In thick 2D material stacks, the continuity of nanoscale percolative pathways becomes critical; otherwise, disconnected regions can prevent precursor access to the substrate, while overstacked regions can block nucleation entirely and disrupt lateral overgrowth. Thus, although solution-based coating provides a scalable and lithography-free route to mask formation, achieving reliable epitaxial connectedness through thick, polydisperse 2D material assemblies remains a key challenge.

In this work, we demonstrate that these challenges can be overcome by employing solution-processed h-BN layers as a self-adjustable mask stack for thru-hole epitaxy. The h-BN fragments, spin-coated onto the substrate, form a loosely stacked network that remains spontaneously rearrangeable during growth. As Ga and N precursors diffuse through and around these fragments, subtle self-adjustment of the stack occurs, expanding nanoscale percolative pathways and enabling stable epitaxial bonding to the substrate.  In this context, we define the term ''self-adjusting'' to describe the ability of loosely stacked h-BN fragments to undergo localized structural reconfiguration—such as shifting, lifting, or reorienting—in response to GaN growth. This behavior enables dynamic modulation of percolation pathways during epitaxy and contributes to both vertical accessibility and dislocation blocking. For consistency, we refer to this dynamic rearrangement as “self-adjusting” throughout the manuscript.  Remarkably, despite the substantial overall thickness—on the order of several hundred nanometers—the self-adjustable, percolatively connected h-BN mask facilitates both crystallographic alignment and defect-suppressed GaN nucleation. This spontaneous, self-adjusting mask behavior provides a scalable and effective alternative to conventional lithographically patterned or rigidly transferred 2D material masks. Our approach demonstrates that robust epitaxial connectedness can be reliably established through thick, polydisperse 2D material assemblies—paving the way for solution-processable selective-area epitaxy.


\section{Experimental}
$h$-BN powder with an initial lateral size of approximately 1~$\mu$m was thermally treated in KOH at 120$^{\circ}$C to promote exfoliation, followed by repeated rinsing with deionized (DI) water using vacuum filtration. The resulting material was then subjected to sonication to further reduce the lateral size of the $h$-BN fragments. After sonication, the solution was evaporated to dryness, and the remaining $h$-BN flakes were collected. These flakes were subsequently redispersed in ethanol to a concentration of 0.0756~g/mL. The suspension, containing $h$-BN fragments of various lateral sizes and thicknesses, was spin-coated onto $c$-plane sapphire substrates at 2400~rpm for two minutes. In this work, the term “fragments” refers to multilayered $h$-BN structures with typical thicknesses on the order of several tens of nanometers, forming loosely stacked networks across the substrate. For the growth of GaN, the flow rates of NH$_3$ and HCl, growth temperature, and growth duration were 1~slm, 10~sccm, 978$^{\circ}$C, and 10~minutes, respectively. Atomic force microscopy (AFM) topography images and optical microscopy images were acquired using a Park Systems XE-100 AFM and an Olympus BX51M metallurgical microscope, respectively, at the Multi-Dimensional Material Convergence Research Center, Kyung Hee University.

\section{Results and discussion}

\begin{figure}
\includegraphics[width=1.0\columnwidth]{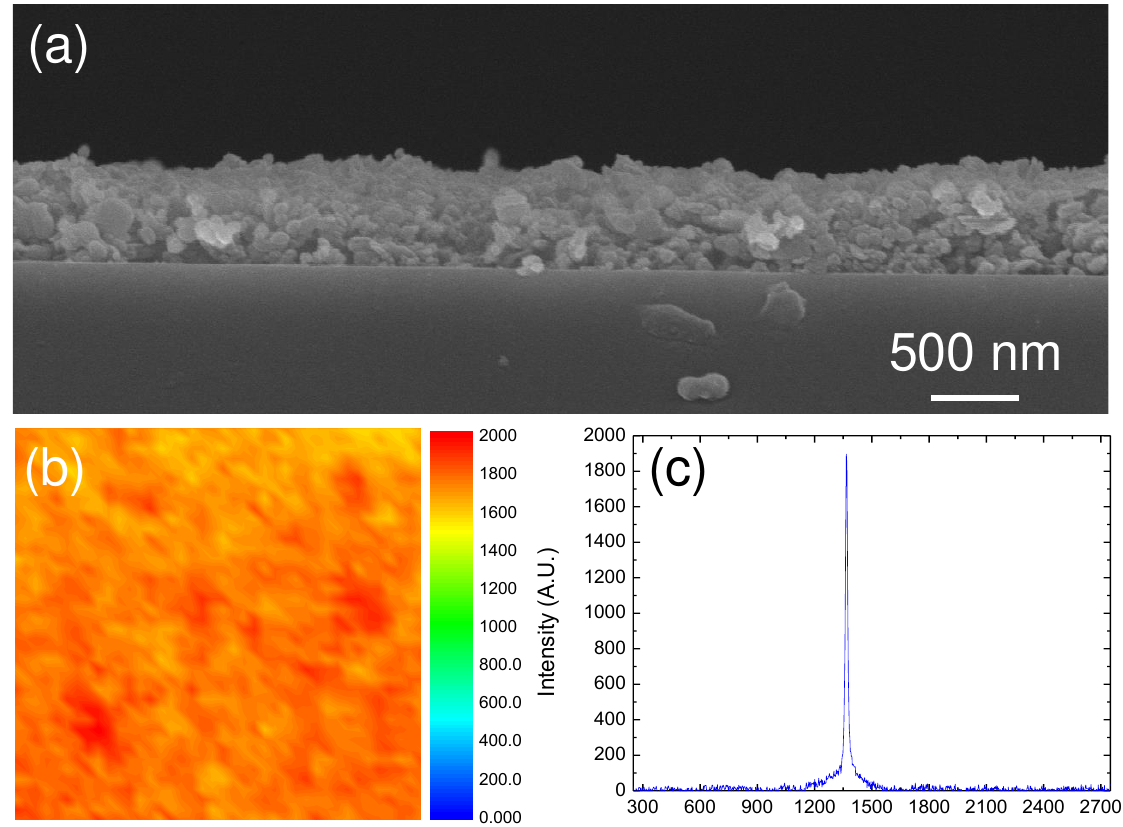}
\caption{(a) Cross-sectional SEM image of $h$-BN fragments spin-coated onto a $c$-plane sapphire substrate.  (b) Raman map of the sample surface over a 20$\times$20$\mu$m$^2$ area and (c) representative Raman spectra, which exhibit a characteristic $h$-BN peak at 1367~cm$^{-1}$, confirming the presence of $h$-BN.  The absence of detectable sapphire-related Raman signals indicates that the $h$-BN fragments provide relatively uniform and sufficiently thick coverage across the mapped region.}
\label{OM-Raman-hBN}
\end{figure}

Figure~\ref{OM-Raman-hBN}(a) shows that the spin-coated $h$-BN layer on the $c$-plane sapphire substrate has an estimated thickness of several hundred nanometers. Opaque $h$-BN fragments are scattered across the surface, as seen in the optical microscopy image. 
Raman spectra collected from the fragments [Figs.~\ref{OM-Raman-hBN}(b)--(c)] exhibit a characteristic peak at approximately 1370~cm$^{-1}$, confirming the presence of $h$-BN.
Raman mapping further reveals relatively uniform fragment coverage across the surface, with the fractional area coverage estimated to be close to unity.

Importantly, due to the irregular shapes and lateral sizes of the $h$-BN fragments—typically on the order of a few hundred nanometers—perfect alignment during spin-coating is inherently constrained.  This imperfect stacking leads to the formation of interconnected nanoscale pathways between adjacent fragments, enabling percolative transport.

To promote GaN nucleation through percolative pathways on $c$-plane sapphire areas and subsequent lateral overgrowth over the $h$-BN mask, the density of $h$-BN fragments and the spin-coating conditions were carefully optimized.  These adjustments ensured that sufficient open areas and percolative pathways remained accessible for precursor diffusion and nucleation during subsequent GaN growth.

\begin{figure*}
\includegraphics[width=2.0\columnwidth]{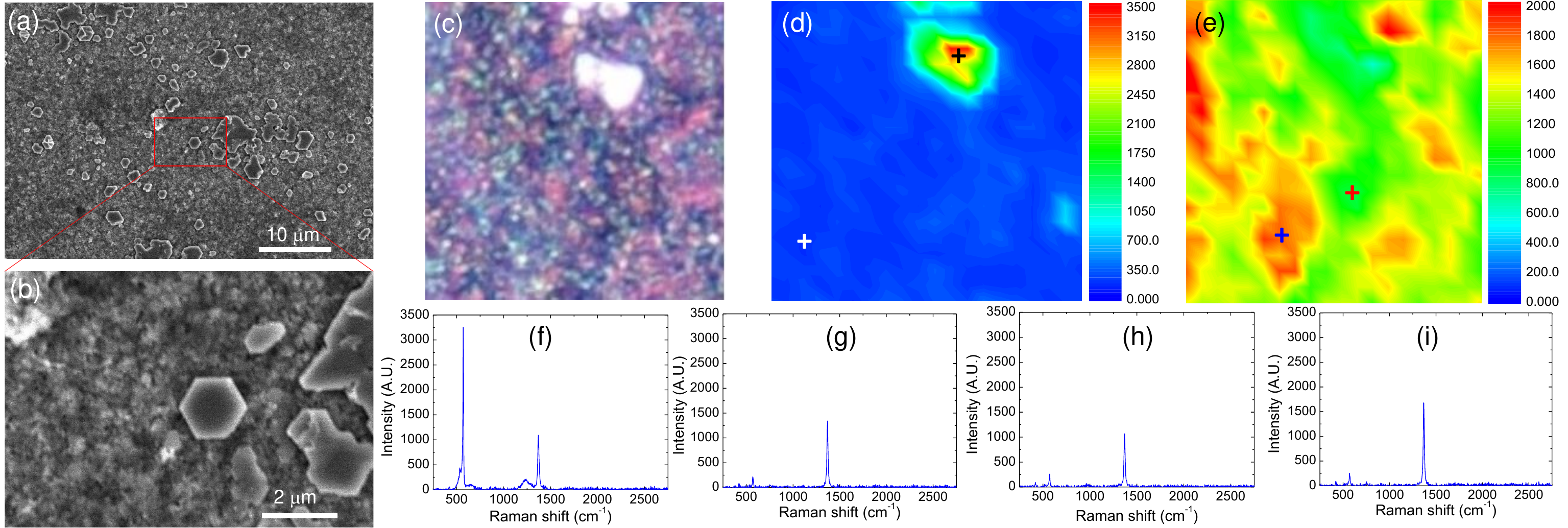}
\caption{(a) and (b) Plan-view SEM images, (c) optical microscopy image of the region used for Raman measurement, Raman mapping of (d) GaN ($\sim$569~cm$^{-1}$) and (e) $h$-BN ($\sim$1369~cm$^{-1}$), (f)-(i) Raman spectra of the region marked by black, white, red, blue cross marks.}
\label{SEM-Raman_after-growth}
\end{figure*}

After confirming $h$-BN coverage, GaN was grown directly on the $h$-BN-coated $c$-plane sapphire substrate without a low-temperature GaN buffer layer. As shown in Fig.~\ref{SEM-Raman_after-growth}, the $h$-BN fragments remained largely intact even after the GaN growth process. Under these conditions, only a small number of isolated GaN domains were observed; importantly, however, their in-plane and out-of-plane crystallographic orientations were well aligned with one another. The observed in-plane alignment of the isolated GaN domains suggests that Ga and N atomic species, supplied over the top surface of the stacked $h$-BN fragments, diffused through several-hundred-nanometer-long percolation pathways and nucleated directly on the underlying sapphire substrate. Despite the fact that the substrate was largely covered by the thick $h$-BN stack, nucleation sites remained accessible through these nanoscale transport channels. From the plan-view SEM images, it can be inferred that nucleation likely occurred at substrate regions that were accessible through local percolation pathways beneath the isolated GaN domains. Conversely, regions devoid of visible GaN domains are thought to be areas where atomic species could not effectively reach the substrate. Intriguingly, however, Raman spectra reveal the presence of GaN-related peaks even in regions where no isolated GaN domains are optically apparent. This unexpected observation is clarified by high-resolution TEM analysis, which reveals additional growth phenomena beyond what is evident from SEM imaging alone.

\begin{figure}[b] 
\includegraphics[width=1.0\columnwidth]{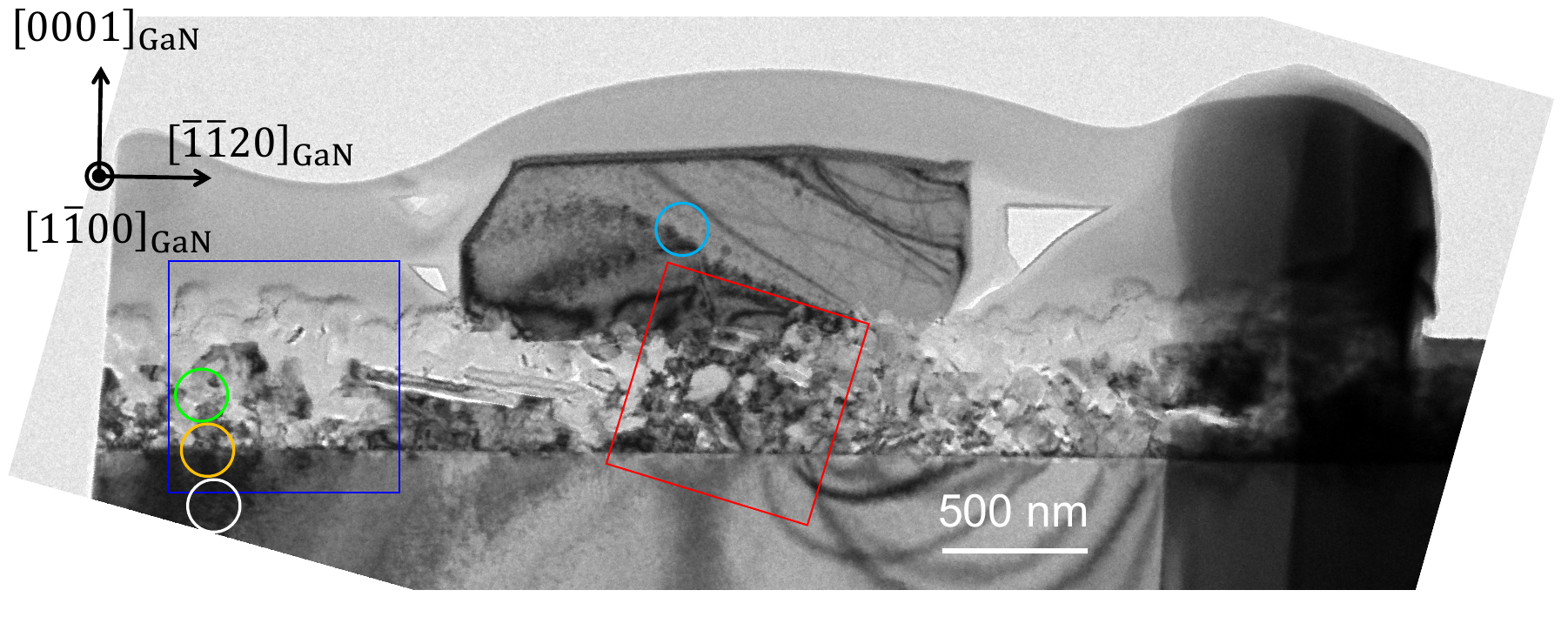}
\caption{Cross-sectional transmission electron microscopy (TEM) image of GaN grown on a spin-coated $h$-BN/$c$-plane sapphire substrate. Regions selected for energy-dispersive spectroscopy (EDS) analysis are indicated by blue and red squares, while areas selected for selected-area electron diffraction (SAED) are marked by cyan, green, orange, and white circles.}
\label{TEM}
\end{figure}

As shown in Fig.~\ref{TEM}, high-resolution transmission electron microscopy (TEM) reveals the presence of a several-hundred-nanometer-thick inhomogeneous layer between the $c$-plane sapphire substrate and both the regions beneath isolated GaN domains and the areas where no GaN domains are visibly present. This cross-sectional observation suggests that (i) GaN nucleation was not confined to isolated spots but rather initiated broadly across the substrate, and (ii) during subsequent growth, the $h$-BN fragments spontaneously self-adjusted in their lateral positions and orientations, leading to a redistribution of stacking thickness across the substrate surface.

The first feature suggests that interconnected percolation pathways enabled atomic species to diffuse through the $h$-BN stack and nucleate epitaxially on the underlying substrate far more effectively than might be expected for such a thick mask. If the $h$-BN fragments had been tightly bound, without the ability to spontaneously self-adjust during GaN growth, percolation pathways would not have formed across the full stack thickness. In that case, nucleation would have been limited to sparse or defect-driven openings.

However, high-resolution TEM images reveal that GaN nucleation occurred almost everywhere on the substrate—not only beneath the isolated GaN domains but also in regions without visible surface domains—suggesting widespread substrate accessibility through percolation. Despite this, the appearance of only a limited number of isolated GaN domains implies that epitaxial growth rates varied significantly across the substrate. Localized self-adjustment of the $h$-BN fragments likely facilitated more efficient pathway formation in regions where isolated GaN domains eventually emerged.

Notably, areas without exposed GaN domains show thick $h$-BN fragment stacking remaining above the composite-like GaN layer, suggesting that upward propagation of GaN through the stack was incomplete in these regions. Thus, GaN nucleated almost uniformly wherever the substrate was accessible, but subsequent vertical growth through the percolative network determined whether isolated, faceted GaN domains could fully emerge above the $h$-BN surface. Where growth successfully breached the $h$-BN stack, isolated GaN domains with well-developed side facets are observed; where it did not, GaN remained buried below.

Overall, the high-resolution TEM observations confirm that epitaxial connectedness—the continuous crystallographic relationship between GaN and the sapphire substrate—was reliably established via self-adjustable percolation pathways formed within the loosely stacked $h$-BN fragments, enabling both nucleation and crystallographic alignment, even through a thick, inhomogeneous barrier layer.

Another important feature revealed by this work is the spontaneous self-adjustment of $h$-BN fragments during GaN growth. While the overall growth mechanism follows the principles of thru-hole epitaxy (THE)—where nucleation occurs through nanoscale percolative pathways rather than explicitly patterned mask openings—the unique behavior of the self-adjusting $h$-BN stack provides additional benefits. During growth, the loosely stacked $h$-BN fragments subtly rearrange, leading to the spontaneous formation of a multilayered masking structure.

This self-adjusting multilayer mask enhances the suppression of threading dislocations (TDs) compared to conventional THE processes, where rigid masks offer less opportunity for dynamic dislocation blocking. The behavior observed here shares conceptual similarities with multi-fold epitaxial lateral overgrowth (ELOG) techniques reported for GaN\cite{Nagahama-JJAP-39-L647,Yoo-SR-7-9663,Gibart-RPP-67-667} and Ga$_2$O$_3$\cite{Kawara-APEX-13-075507}, in which sequential masking suppresses TD propagation. However, in this study, such multilayer masking arises spontaneously, without the need for repeated lithographic steps.

While direct measurement of threading dislocation density (TDD) was not performed in this study due to the limited sampling area inherent to high-resolution TEM, which only captures nanoscale regions and is not suitable for statistically evaluating dislocation densities across the full sample, several indirect observations support the interpretation that dislocation propagation was significantly suppressed. First, the emergence of well-faceted, isolated GaN domains suggests a reduction in dislocation-mediated strain relaxation, as such faceting is typically more pronounced in low-defect-density regions. Second, the absence of vertical threading-like contrast in TEM images taken beneath isolated domains, combined with the preservation of clear crystallographic alignment in SAED patterns, further suggests coherent growth with minimal dislocation insertion. Finally, the observed multilayer-like self-adjusting behavior of the h-BN stack resembles sequential masking strategies previously reported for dislocation filtering, such as multiple ELOG or nanoparticle-based selective growth. Taken together, these factors point to effective suppression of dislocation propagation, even in the absence of direct TDD quantification.

Similar strategies for defect reduction have been reported using Si-N nanoparticle masks formed during growth,\cite{Tanaka-JJAP-39-831} double SiO$_2$ nanopillar masking for $a$-plane InGaN/GaN multiple quantum wells,\cite{Son-JJAP-53-05FL01} and SiN$_x$ interlayers.\cite{Yun-JAP-98-123502} Unlike multiple-ELOG processes that depend on graphene decomposition to open additional nucleation sites,\cite{Lee-JAC-53-1502} the self-adjusting $h$-BN stack in this work remains structurally intact, enabling epitaxial connectedness through dynamically evolving percolative pathways. This feature allows not only spatially confined nucleation—characteristic of THE—but also more effective blocking of dislocation propagation during GaN growth.

\begin{figure}    
\includegraphics[width=1.0\columnwidth]{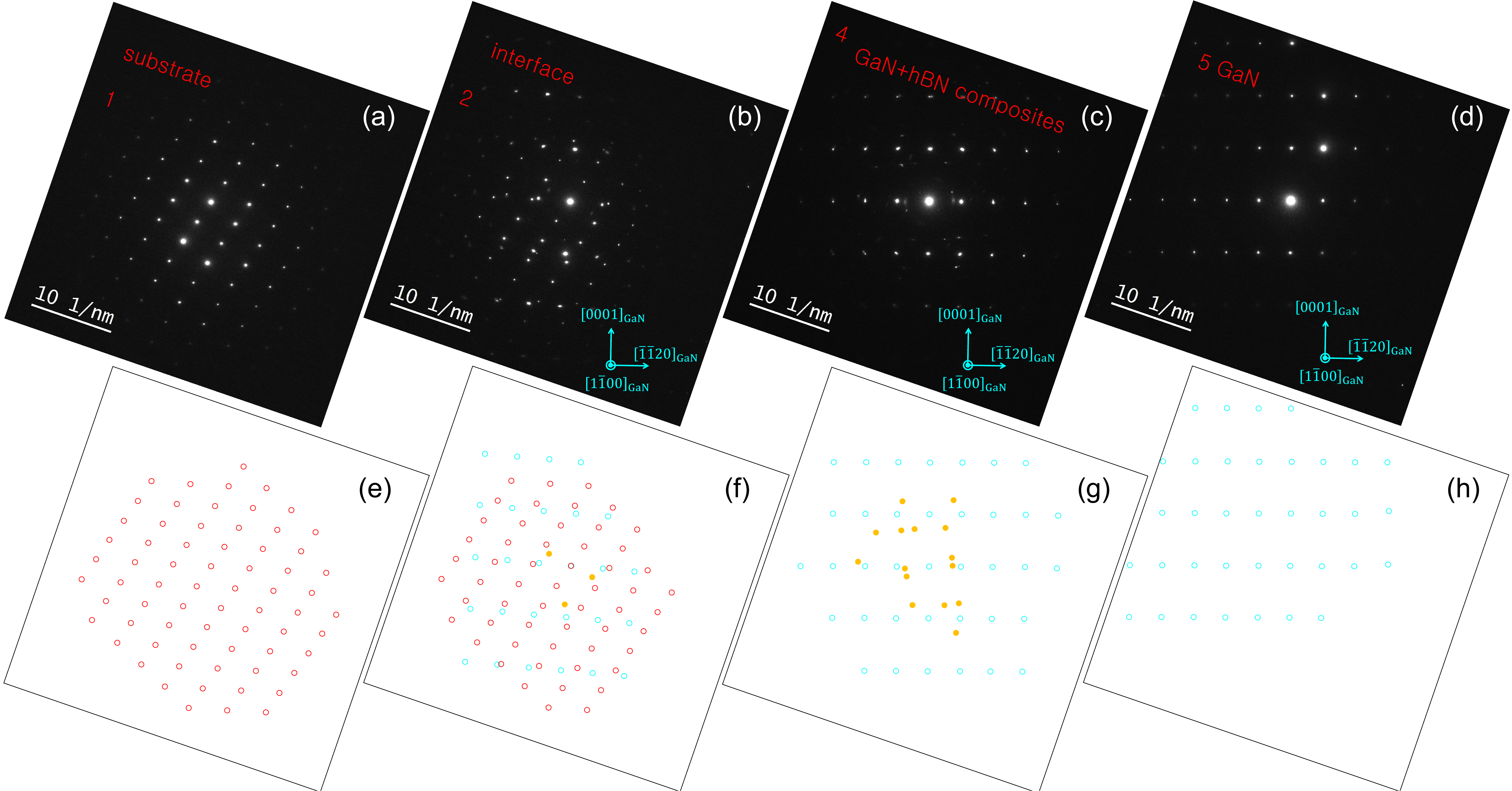}
\caption{Selected area electron diffraction (SAED) patterns obtained from (a) the sapphire substrate (white-circled region), (b) the GaN/sapphire interfacial region (orange-circled region), (c) the composite region containing GaN and self-adjusted $h$-BN fragments (green-circled region), and (d) the pure GaN domain (cyan-circled region), as indicated in Fig.~\ref{TEM}.}
\label{SAED}
\end{figure}

To confirm that the crystallographic alignment of GaN domains with the underlying sapphire substrate was maintained through the percolative pathways, selected area electron diffraction (SAED) measurements were performed at various depths relative to the substrate surface. SAED patterns were obtained from different regions, marked by colored circles in Fig.~\ref{TEM}. As shown in Fig.~\ref{SAED}, the crystallographic orientations of GaN in the interfacial region (yellow-circled area), the composite region containing $h$-BN fragments and GaN (green-circled area), and the pure GaN domain region (cyan-circled area) were found to be identical.

Importantly, diffraction spots corresponding to neither sapphire nor GaN were also observed. These additional spots are attributed to the randomly oriented, self-adjusted $h$-BN fragments embedded within the GaN matrix. The preservation of crystallographic alignment across different vertical positions strongly supports that epitaxial connectedness was successfully established via percolative pathways through the self-adjusting $h$-BN stack.

\begin{figure}
\includegraphics[width=1.0\columnwidth]{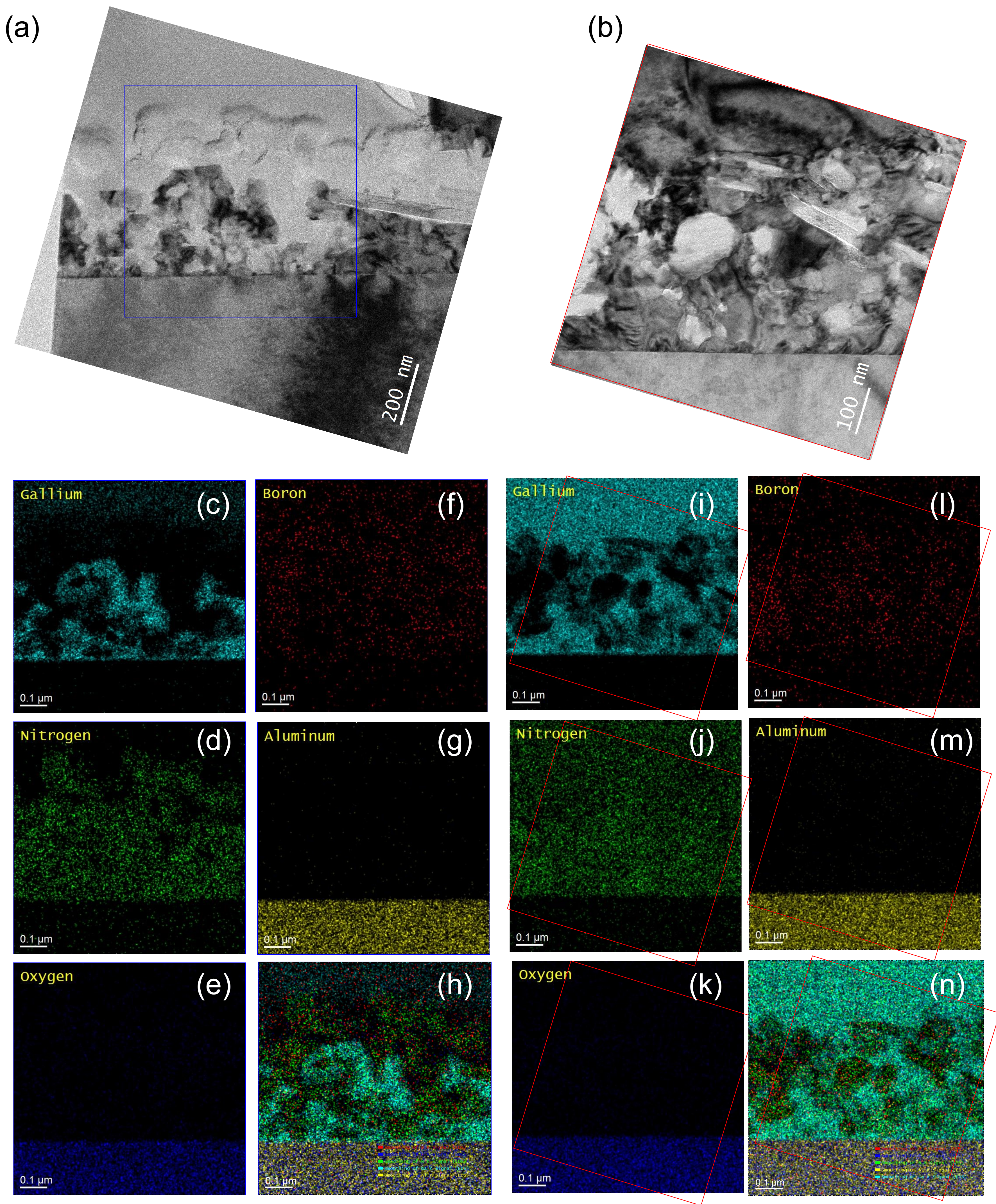}
\caption{High-resolution TEM and EDS mapping of the composite layer formed beneath and around GaN domains on h-BN-coated sapphire substrates.  (a, b) High-resolution TEM images from two different regions: (a) a region without a micron-scale, faceted GaN domain, and (b) a region with a well-defined, isolated GaN domain.  (c–h) EDS elemental maps of the area in (a), showing distributions of Ga (c), N (d), O (e), B (f), Al (g), and a composite overlay of all elements (h).  (i–n) EDS elemental maps of the area in (b), showing corresponding distributions of Ga (i), N (j), O (k), B (l), Al (m), and the composite overlay (n).  These maps confirm that Ga and N are present even beneath regions without exposed GaN domains, and that the composite layer includes both GaN and embedded h-BN fragments.}
\label{EDS}
\end{figure}

Additionally, depth-resolved elemental analysis was performed using energy-dispersive X-ray spectroscopy (EDS). As shown in Fig.~\ref{EDS}(a) and (b), which compare a region without an isolated GaN domain and one with a well-developed domain, respectively, magnified high-resolution TEM images of the regions marked by blue and red squares in Fig.~\ref{TEM} reveal that a composite layer is present across the entire substrate surface, regardless of whether a micron-scale single-crystalline GaN domain was formed above.  EDS elemental mapping confirms that this composite layer is composed primarily of GaN and BN. These results suggest that Ga and N species diffuse through nanoscale voids between $h$-BN fragments and nucleate directly on the underlying sapphire substrate. As nucleation proceeds, the growing GaN nuclei not only expand laterally but also partially lift and reorient the $h$-BN fragments, enabling a multilayer-like epitaxial lateral overgrowth process.  As GaN growth continues and the composite layer thickens, certain regions experience more rapid vertical growth—eventually exhausting the local $h$-BN fragments—which results in the formation of isolated, well-faceted single-crystalline GaN domains.
Thus, the epitaxial connectedness between GaN and the sapphire substrate, mediated by the spontaneous rearrangement of $h$-BN and percolative transport pathways, is strongly supported by both high-resolution TEM observations and EDS mapping.

A conceptually related phenomenon has been reported in the migration-enhanced encapsulated growth (MEEG) technique, where quasi-freestanding epitaxial graphene formed on the surface of a SiC substrate was observed to lift during GaN intercalative growth between the graphene and SiC.\cite{Al-Balushi-NM-15-1166} In that work, GaN was grown either in a two-dimensional or three-dimensional mode beneath the graphene layer. However, the graphene used was only a few atomic layers thick, and the focus was on intercalative growth rather than lateral overgrowth through or over the 2D material.

In contrast, the present work demonstrates a markedly different regime: the $h$-BN mask is several hundred nanometers thick—approximately a thousand times thicker than the graphene layers used in MEEG—and consists of loosely stacked, self-adjusting fragments. Despite this substantial thickness, epitaxial connectedness was successfully established through dynamic percolative pathways within the $h$-BN stack. Moreover, the self-adjusting multilayer mask behavior observed during GaN overgrowth led to effective suppression of threading dislocations.


Another important issue to address is whether GaN nucleation can occur on sapphire substrates via atomic diffusion through percolation pathways extending over several hundred nanometers. Although the physical structures differ, efficient molecular and atomic transport through nanoscale channels has been demonstrated in previous studies.\cite{Radha-Nature-538-222} Furthermore, long-distance, low-friction molecular transport through two-dimensional confined spaces has been experimentally verified, supporting the feasibility of superlong diffusion across nanoscale pathways.\cite{Geim-NL-21-6356} While the diffusion in this study occurs through percolative networks formed by loosely stacked $h$-BN fragments—rather than through well-defined atomic channels—these prior findings reinforce the plausibility that precursor species could traverse the interconnected voids, reach the substrate surface, and enable GaN nucleation even through thick, disordered, 2D-material-based masks.

The concept of "self-adjustment" in the h-BN stack is supported by several morphological and structural observations. High-resolution TEM reveals that the thickness of the h-BN assembly varies laterally across the sample, and that GaN is present beneath both thick and thin regions of the stack. This suggests that the loosely stacked h-BN fragments are not statically fixed but instead undergo local rearrangement during GaN growth. Such rearrangement likely occurs due to mechanical interactions at the growth front—such as vertical pressure from nucleating GaN or lateral stress from expanding domains—which can displace or reorient adjacent h-BN flakes. In regions where GaN successfully breaches the mask, the overlying h-BN appears to have been locally thinned or lifted, enabling vertical propagation. In contrast, regions with thicker, unadjusted stacking show buried GaN growth, indicating that vertical overgrowth was obstructed. These findings imply that the h-BN stack undergoes localized structural reconfiguration in response to GaN growth dynamics, effectively modulating the accessibility and geometry of percolation pathways in real time.

Building on this understanding of transport phenomena, it is also important to consider the broader implications for device applications. From a device perspective, the miniaturization of micro-light-emitting diode ($\mu$-LED) structures intensifies the need for defect control. As device dimensions shrink, sidewall leakage currents become increasingly problematic,\cite{Gonzalez-ACSPhotonics-10-4031} and the detrimental impact of threading dislocations on device efficiency and reliability becomes more pronounced.\cite{Liu-PhotoniX-5-23,Wang-OE-32-31463} Therefore, minimizing both sidewall damage and dislocation density is essential for high-performance $\mu$-LED applications. The self-adjusting $h$-BN mask approach demonstrated in this work addresses both challenges simultaneously: (i) by introducing a self-adjusting mask layer that effectively blocks threading dislocation propagation, and (ii) by enabling selective nucleation and growth without requiring physical chip delineation processes, thereby mitigating sidewall damage associated with conventional etching steps. This strategy offers a promising pathway for defect-suppressed, scalable fabrication of next-generation $\mu$-LEDs and other high-performance optoelectronic devices.

The use of solution-processed h-BN suspensions offers a promising route toward scalable mask formation for selective-area epitaxy, particularly because it avoids the need for transfer steps or lithographic patterning. While the current study focused on small-area samples suitable for high-resolution analysis, the spin-coating method employed here is inherently compatible with large-area and three-dimensional substrate geometries. However, the uniformity and reproducibility of flake distribution, thickness, and percolative connectivity across wafer-scale areas remain to be systematically evaluated. Future work will explore process optimization strategies—such as surfactant-assisted dispersion, flake size selection, or spin profile tuning—to ensure large-area uniformity and consistent growth outcomes. Nonetheless, the present results demonstrate the feasibility of achieving epitaxial connectedness and defect suppression in locally thick, disordered regions, which is an essential first step toward broader-scale application.

\section{Conclusion}
We have demonstrated that initial nucleation and subsequent lateral growth of GaN can be achieved on a crystalline substrate fully covered with a spin-coated stack of $h$-BN flakes. Despite the substantial thickness and inherent disorder of the stack, the formation of a network of nanoscale percolation pathways enabled robust epitaxial registry between the GaN and the underlying substrate. During growth, the loosely bound $h$-BN fragments spontaneously self-adjusted, expanding the percolation network and stabilizing the epitaxial connection while preserving limited mechanical coupling.

This self-adjustable percolative network not only suppresses threading dislocations but also suggests the potential for facile detachment of the grown film, owing to the minimized mechanical anchoring. Furthermore, by employing solution-processed $h$-BN flakes, this approach enables scalable coating of three-dimensional substrates without requiring transferred or directly grown 2D layers. Overall, this work establishes a scalable and versatile strategy for selective-area epitaxy, providing a new pathway for defect-suppressed, flexible optoelectronic device fabrication. This method provides a foundation for integrating 2D material-based masks in scalable device manufacturing platforms, with potential applicability in micro-LEDs, sensors, and other heteroepitaxial systems.

\section{acknowledgement}
This work was supported by the National Research Foundation of Korea(NRF) grant funded by the Korea government (MSIT) (RS-2021-NR060087, RS-2023-00240724) and through Korea Basic Science Institute (National research Facilities and Equipment Center) grant (2021R1A6C101A437) funded by the Ministry of Education.


%

\end{document}